\pdfoutput=1
\documentclass[preprint,preprintnumbers,amsmath,amssymb,superscriptaddress,colorlinks]{revtex4}
 
\usepackage[bookmarks=false]{hyperref}
\usepackage{graphicx}
\usepackage{epstopdf}
\usepackage{dcolumn}
\usepackage{bm}
\usepackage{color}
\usepackage{ulem}

\begin{document}

\title{ Dijet asymmetry at the Large Hadron Collider}
\author{Clint Young}
\email{clinty@physics.mcgill.ca}
\affiliation{Department of Physics, McGill University, 3600 University Street, Montreal, Quebec, H3A\,2T8, Canada}
\author{Bj\"orn Schenke}
\email{bschenke@quark.phy.bnl.gov}
\affiliation{Physics Department, Bldg. 510A, Brookhaven National Laboratory, Upton, NY 11973, USA}
\author{Sangyong Jeon}
\email{jeon@physics.mcgill.ca}
\affiliation{Department of Physics, McGill University, 3600 University Street, Montreal, Quebec, H3A\,2T8, Canada}
\author{Charles Gale}
\email{gale@physics.mcgill.ca}
\affiliation{Department of Physics, McGill University, 3600 University Street, Montreal, Quebec, H3A\,2T8, Canada}

\date{\today}

\begin{abstract}

The \textsc{martini} numerical simulation allows for direct comparison of theoretical model calculations and the latest results for dijet asymmetry from the ATLAS and CMS collaborations. In this paper, partons are simulated as undergoing radiative and collisional processes throughout the evolution of central lead-lead collisions at the Large Hadron Collider. Using hydrodynamical background evolution determined by a simulation which fits well with the data on charged particle multiplicities from ALICE and a value of $\alpha_s\approx 0.25-0.3$, the dijet asymmetry is found to be consistent with partonic energy loss in a hot, strongly-interacting medium.

\end{abstract}

\maketitle

\section{Introduction}

Within a month of running, the heavy-ion programs at the Large Hadron Collider has produced important results. The anisotropic flow is similar 
to the flow measured at the RHIC experiment and suggests intensive bulk properties comparable to what would also describe lower energy collisions 
\cite{Aamodt:2010pb, Teaney:2001av, Schenke:2011tv}. However, because $\sqrt{s_{NN}}=2.76\;{\rm TeV}$ at the LHC, significantly 
larger than the center-of-mass energies achieved at the RHIC, far more energetic jets are kinematically accessible at the LHC. The ATLAS 
collaboration was able to measure over 1000 dijets where the leading jet has transverse energy $E_T > 100\;{\rm GeV}$ and the opposing 
jet has energy $E_T > 25\;{\rm GeV}$ \cite{Collaboration:2010bu}. The CMS collaboration performed a similar analysis on a large sample of 
jets ($E_{T1}>120\;{\rm GeV}$, $E_{T2}>50\;{\rm GeV}$). These results are a significant improvement over the results from the RHIC, where 
the total energies of the jets were far lower and therefore harder to separate from fluctuations in the underlying bulk. Also, the models for 
partonic evolution rely on the probe parton having high energy, and when this separation of energies exists one can expect the hadronization of 
these partons to be described well with vacuum fragmentation functions. Several theoretical studies of these results are now available: Majumder 
 and Che et al. examined the supression of high-$p_T$ hadrons assuming purely radiative energy loss and found good agreement with the rising 
$R_{AA}$ for high transverse momentum seen in the latest analysis of ALICE \cite{Majumder:2011uk, Che:2011vt}.
Qin and M\"uller studied the evolution of the whole jet shower, as the jet propagates through the quark-gluon plasma and interacts with the 
medium \cite{Qin:2010mn}. Casalderrey-Solana et al.  
\cite{CasalderreySolana:2010eh} conclude that the removal of soft components from within the jet cone will induce a dijet asymmetry.
Also, Lokhtin {\it et al.} use the \textsc{pyquen} model to quantify the ``jet-trimmingÓ \cite{Lokhtin:2011qq}.

In this paper, we apply \textsc{martini} to the lead-lead collisions at the LHC \cite{Schenke:2009gb, Schenke:2009vr}. In 
Section \ref{transport}, the physics behind \textsc{martini} is reviewed, as well as the description of the bulk of 
heavy-ion collisions with 3+1-dimensional hydrodynamics. In Section \ref{results}, runs of \textsc{martini} with cuts given by the ATLAS and 
CMS detectors and their analyses are compared with the experimental results for both $dN/dA_J$ (the yield of dijets differential in $A_J$, 
where $A_J$ measures the energy anisotropy of dijets) and $dN/d\phi$. In Section \ref{conclusions}, we briefly discuss these results and 
their relationship with other experimental results at the LHC.

\section{Transport of high-energy partons and \textsc{martini}}
\label{transport}

\textsc{martini} solves the rate equations
\begin{eqnarray*}
\frac{dP_{q\bar{q}}(p)}{dt} &=& \int_k P_{q\bar{q}}(p+k) \frac{d\Gamma^q_{qg}(p+k,k)}{dkdt}-P_{q\bar{q}}(p)\frac{d\Gamma^q_{qg}(p,k)}{dkdt}+P_g(p+k)
\frac{d\Gamma^g_{q\bar{q}}(p+k,k)}{dkdt}{\rm ,} \\
\frac{dP_g(p)}{dt} &=& \int_k P_{q\bar{q}}(p+k)\frac{d\Gamma^q_{qg}(p+k,p)}{dkdt}+P_g(p+k)\frac{d\Gamma^g_{gg}(p+k,k)}{dkdt} \\
 & & -P_g(p)
\left( \frac{d\Gamma^g_{q\bar{q}}(p,k)}{dkdt}+\frac{d\Gamma^g_{gg}(p,k)}{dkdt}\Theta(2k-p) \right){\rm ,} \\
\end{eqnarray*}
where the various {\it differential} rates $d\Gamma^i_{lm}(p,k)/dkdt$ determine the splitting of partons $l$ and $m$, one with momentum $k$, from 
a parton $i$ with momentum $p$ \cite{Turbide:2005fk}.

In its current implementation, \textsc{martini} uses rates which take into account both radiative and collisional QCD processes, calculated at finite temperature.
Collisional processes involve soft momentum transfers sensitive to the gluon's screening mass and 
therefore, hard thermal loop results at leading order are used to describe these processes. 
For the elastic processes \textsc{martini} does not depend on the ``diffusive approximation'': there is no need to assume that the rates are 
only significant when $\omega$ is small \cite{Schenke:2009ik}.
Radiative energy loss is 
modeled using the Arnold-Moore-Yaffe approach, where the interference of bremsstrahlung gluons from multiple scatterings is taken into 
account with an LPM-like integral equation for the energy loss rate \cite{Arnold:2001ms, Arnold:2001ba, Arnold:2002ja}. 

The momenta of high-energy partons are sampled using \textsc{pythia} event generation \cite{Sjostrand:2006za}, and their initial positions in the transverse 
plane of heavy-ion collisions are sampled according to $n_B(x, y, b)$, the distribution of binary collisions for a given impact parameter $b$ of the collision. 
These partons are then evolved through the background of bulk particles. For the results of Section \ref{results}, this evolving background is modeled 
using \textsc{music}, a 3+1-dimensional hydrodynamic simulation \cite{Schenke:2010nt}. The use of 3+1-dimensional hydrodynamics allows jets at different 
rapidities to evolve differently, as one should expect.

For the results in Section \ref{results}, \textsc{martini} is run with $\alpha_s=0.25{\rm ,}\; 0.27{\rm ,}\;{\rm and}\;0.3$ including both collisional and 
radiative processes. The finite-temperature rates for these processes are determined by the temperatures and flow in lead-lead collisions as 
simulated with \textsc{music} for an impact parameter of $b=2.31\;{\rm fm}$, reproducing the multiplicities of the 0-10\% centrality class.
In this study we use a simulation with ideal hydrodynamics starting with the averaged initial conditions.

In summary, the strengths of \textsc{martini} include the inclusion of combined radiative and elastic processes, its 
integration with \textsc{pythia} and Glauber model calculations for both sampling of the initial parton distributions in momentum and 
position and the fragmentation of the evolved partons into hadrons, and the ability to evolve the partons in a background medium obtained 
from realistic hydrodynamical simulations.

\section{Results for lead-lead collisions measured at ATLAS and CMS}
\label{results}

Once high-energy partons have evolved and hadronized, the resulting hadrons must then be reconstructed into jets. For the best possible comparison with the results of the LHC, we use the same anti-$k_t$ jet reconstruction that the ATLAS collaboration uses \cite{Cacciari:2008gp}. These algorithms 
depend on the definition of distances between two 4-momenta:
\begin{equation}
d_{ij}={\rm min} \left( \frac{1}{k^2_{it}}, \frac{1}{k^2_{jt}} \right) \frac{(\phi_i-\phi_j)^2+(y_i-y_j)^2}{R^2}{\rm .}
\end{equation}
The distances are determined between all pairs of final-state particles whose energies are large enough to trigger the calorimeters, and starting with the smallest distance, 4-momenta close to each other are clustered and added together and final jets are determined. The implementation of this algorithm that we used is \textsc{fastjet}, publicly available online \cite{fastjet}.

Once the clustering of hadrons into jets is complete, the jet with highest $E_T$ is determined, and the highest energy jet whose azimuthal angle from the leading jet $\Delta \phi > \pi/2$ (or $2\pi/3$, as is the case with the CMS analysis) is also determined. If the energies of this dijet are high enough to make it into the given detector's analysis, they are recorded and binned.
 
\begin{figure}
\includegraphics[height=9cm]{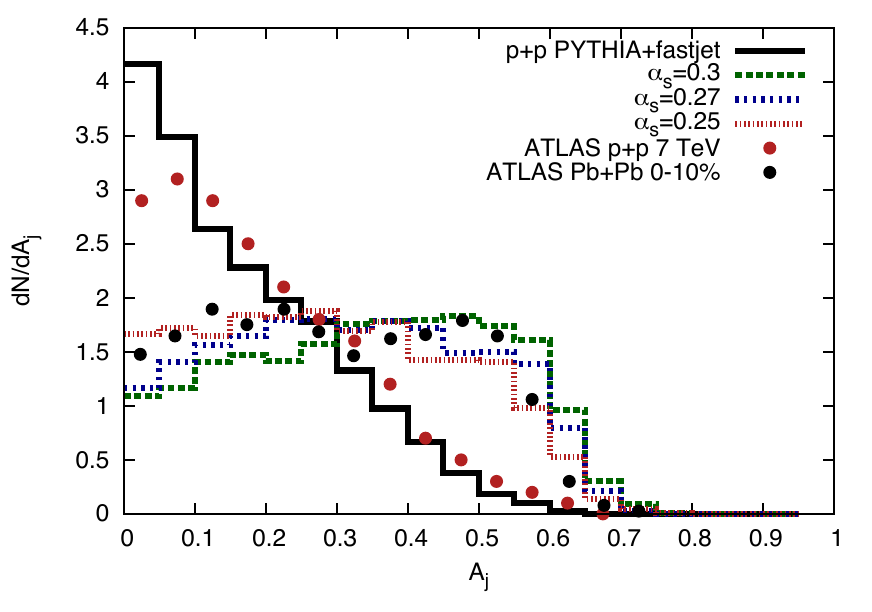}
\caption{
The differential yield $dN/dA_J$ for proton-proton collisions at (solid) and 
for lead-lead collisions (dashed, dotted), both at $\sqrt{s}=2.76\;{\rm GeV}$ for each nucleon-nucleon collision.
}
\label{dNdA_ET07}
\end{figure}

\begin{figure}
\includegraphics[height=9cm]{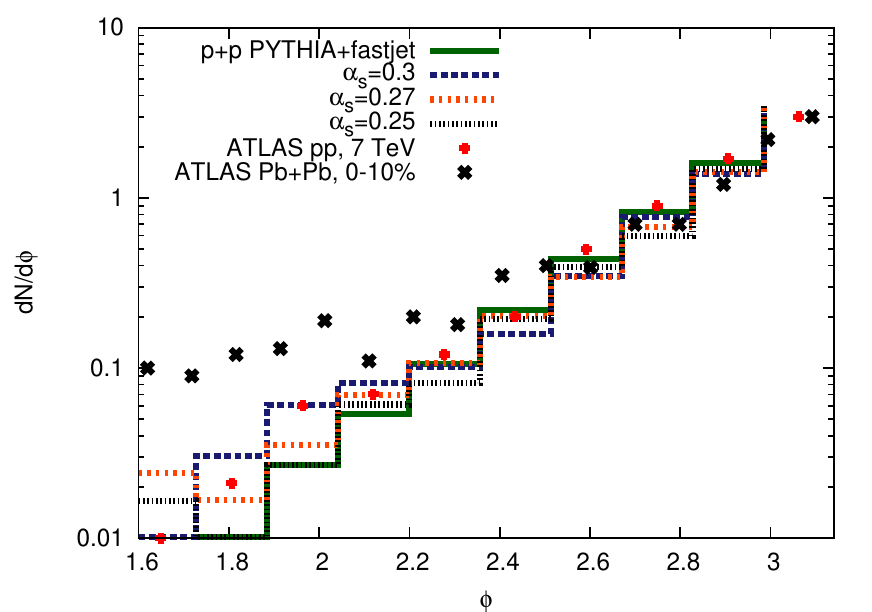}
\caption{
The differential yield $dN/d\phi$ for proton-proton collisions at $\sqrt{s}=2.76\;{\rm GeV}$ (solid) and for 
lead-lead collisions (dashed, dotted).
 }
\label{dNdphi_ET07}
\end{figure}

In Figure \ref{dNdA_ET07}, we show the results for ATLAS, in the 0-10\% centrality range, for the differential yield $dN/dA_J$, where $A_J=\frac{E_{T1}-E_{T2}}{E_{T1}+E_{T2}}$ is a measure of the transverse energy asymmetry of the dijets. The ATLAS results used are from the 
latest analysis using $R=0.4$ \cite{Cole:QM2011}; there was little dependence of $R$ found in the latest results, suggesting partonic energy loss as the dominant mechanism leading to dijet asymmetry. Our results are compared with p+p events using \textsc{pythia} and \textsc{fastjet}, and the differential yields are normalized to one. In Figure \ref{dNdphi_ET07}, we show the differential yields $dN/d\phi$, where $\phi$ is the azimuthal opening angle for the dijets.

The results are shown for both $\alpha_s=0.25{\rm,}\; 0.27{\rm ,}\; {\rm and}\; 0.3$. From examining the results for $dN/d\phi$ compared 
with ATLAS, it is clear that the \textsc{martini} model constrains tightly the only parameter in the model, $\alpha_s$.  

On the other hand, Fig. \ref{dNdphi_ET07} shows no significant difference in the distribution of dijets between proton-proton and lead-lead collisions. The 
experimental results show a significant increase in the yield at small $\phi$ in lead-lead collisions over what was observed in 
proton-proton collisions. This enhancement, while significant, affects a relatively small number of dijets 
in ATLAS' sample, and could be due to complications facing jet reconstruction in heavy-ion collisions 
with fluctuating soft backgrounds. This possible explanation was demonstrated recently by Cacciari, Salam, and Soyez, without any consideration of jet quenching \cite{Cacciari:2011tm}. We are currently working on including the event-by event fluctuations of the initial conditions to take this effect into account. However
we should point out that these fluctuations affect a relatively small number of jets and does not significantly affect our results besides the differential 
yield at small angles (which is clear when plotted semi-logarithmically).

\begin{figure}
\includegraphics[height=9cm]{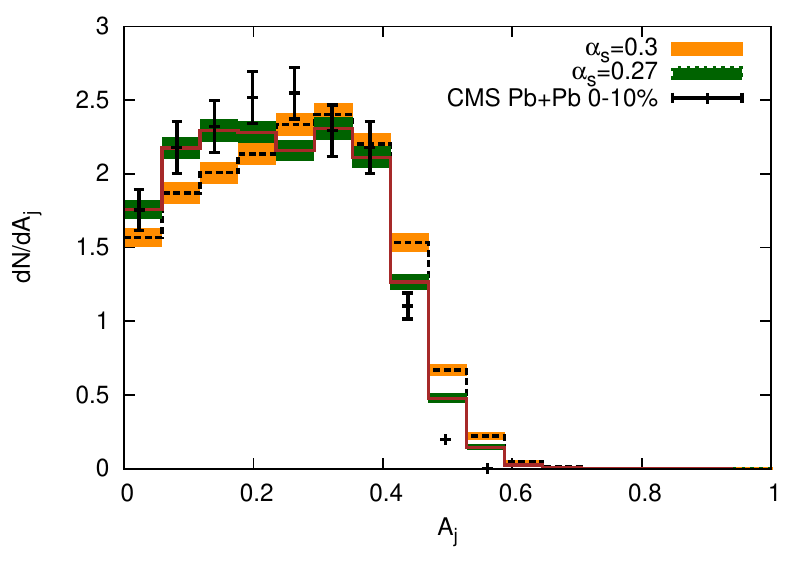}
\caption{
(Color online)  The differential yield $dN/dA_J$ for lead-lead collisions at $\sqrt{s}=2.76\;{\rm GeV}$.
}
\label{dNdA_CMS}
\end{figure}

Finally, Figure \ref{dNdA_CMS} shows the differential yield in $A_j$ determined by CMS' dijet sample, compared with \textsc{martini}'s results based on 
CMS' kinematical cuts. 

\section{Conclusions}
\label{conclusions}

The study reported here utilizes
the pQCD and thermal-QCD based MARTINI numerical simulation with a hydrodynamic background determined by \textsc{music} and 
full jet reconstruction using \textsc{fastjet}.
Using only one free parameter - $\alpha_s$ - we can explain a large part of the jet asymmetries observed
in the recent ATLAS and CMS experiments at the LHC as the consequences of high energy jets interacting with
the evolving QGP medium. According to a recent study, the discrepancy in the angular asymmetry may be due
to the fluctuating soft background. We are currently accumulating a statistically significant number of event-by-event
hydrodynamics events to study this effect further. For $dN/dA_J$, it is shown that our approach
describes the CMS data significantly better than the relatively softer ATLAS data. This may be again due to fact
that compared to the averaged initial conditions, the event-by-event initial conditions have significant granularity.
It is conceivable that the more localized hot spots in the fluctuating case affects the shape of $dN/dA_J$ more
for the relatively softer partons since most energy loss occurs early in the evolution.
These and other effects such as the viscosity of the medium are currently under investigation.

\section{Acknowledgments}

CY thanks Jean Barrette, Vasile Topor Pop, and Todd Springer for useful discussions. CG, SJ, and CY were supported by the Natural Sciences 
and Engineering Research Council of Canada and BPS was supported in part by the US Department of Energy under DOE Contract No. 
DEAC02-98CH10886 and by a Lab Directed Research and Development Grant from Brookhaven Science Associates.

\end{document}